\documentclass[sigconf]{acmart}

\usepackage{graphicx}
\usepackage{textcomp}
\usepackage{xcolor}
\usepackage{enumitem}
\usepackage[ruled,vlined]{algorithm2e}
\usepackage{pifont}
\usepackage{subcaption}
\usepackage{multirow}
\usepackage{booktabs}

\newcommand{\tang}[1]{\textcolor{black}{#1}}
\newcommand{\blue}[1]{\textcolor{black}{#1}}

\newcommand{\black}[1]{\textcolor{black}{#1}}

\newcommand{\pmc}[0]{{\emph{PMC$^2$}}}
\newcommand{\jpgOnMobileName}[0]{{``JPEG Compression on Mobile''}}

\newcounter{YaoNumberOfComments}
\stepcounter{YaoNumberOfComments}

\AtBeginDocument{%
  \providecommand\BibTeX{{%
    \normalfont B\kern-0.5em{\scshape i\kern-0.25em b}\kern-0.8em\TeX}}}

\renewcommand\footnotetextcopyrightpermission[1]{}

\begin{document}

\title{Privacy-Preserving Multimedia Mobile Cloud Computing Using Protective Perturbation}


\author{Zhongze Tang}
\affiliation{%
  \institution{Rutgers University}
  \city{Piscataway}
  \state{NJ}
  \country{USA}
}
\email{zhongze.tang@rutgers.edu}

\author{Mengmei Ye}
\affiliation{%
  \institution{IBM Research}
  \city{Yorktown Heights}
  \state{NY}
  \country{USA}
}
\email{mye@ibm.com}

\author{Yao Liu}
\affiliation{%
  \institution{Rutgers University}
  \city{Piscataway}
  \state{NJ}
  \country{USA}
}
\email{yao.liu@rutgers.edu}

\author{Sheng Wei}
\affiliation{%
  \institution{Rutgers University}
  \city{Piscataway}
  \state{NJ}
  \country{USA}
}
\email{sheng.wei@rutgers.edu}

\begin{abstract}
Mobile cloud computing has been adopted in many multimedia applications, where the resource-constrained mobile device sends multimedia data (e.g., images) to remote cloud servers to request computation-intensive multimedia services (e.g., image recognition). While significantly improving the performance of the mobile applications, the cloud-based mechanism often causes privacy concerns as the multimedia data and services are offloaded from the trusted user device to untrusted cloud servers. Several recent studies have proposed perturbation-based privacy preserving mechanisms, which obfuscate the offloaded multimedia data to eliminate privacy exposures without affecting the functionality of the remote multimedia services. However, the existing privacy protection approaches require the deployment of computation-intensive perturbation generation on the resource-constrained mobile devices. Also, the obfuscated images are typically not compliant with the standard image compression algorithms and suffer from significant bandwidth consumption. In this paper, we develop a novel privacy-preserving multimedia mobile cloud computing framework, namely \pmc{}, to address the resource and bandwidth challenges. \pmc{}{} employs secure confidential computing in the cloud to deploy the perturbation generator, which addresses the resource challenge while maintaining the privacy. Furthermore, we develop a neural compressor specifically trained to compress the perturbed images in order to address the bandwidth challenge. We implement \pmc{}{} in an end-to-end mobile cloud computing system, based on which our evaluations demonstrate superior latency, power efficiency, and bandwidth consumption achieved by \pmc{} while maintaining high accuracy in the target multimedia service.
\end{abstract}

\maketitle
\pagestyle{plain}

\section{Introduction}
Mobile cloud computing has been adopted in many mobile multimedia applications to compensate for the limited computing resources on the mobile devices~\cite{Wang-TM13,Cer-ICNC14,Xu-WirelessComm13,Khan-Access22,Alh-WWW22,Zhao-TMC22,Chen-VehicularComm17,Tian-CNS17}. In a typical mobile multimedia cloud computing system, the mobile device sends input data (e.g., images) to a remote cloud server, which provides computation-intensive services (e.g., image recognition) that cannot be deployed on the mobile device due to resource limitations or intellectual property regulations. With the rapid advancements of high-speed communication networks and high-capacity data centers, mobile cloud computing has demonstrated significant performance advantages over traditional mobile-only computing in many application domains, such as image recognition/annotation~\cite{Tian-CNS17}, object detection~\cite{Chen-VehicularComm17}, and gaming~\cite{Alh-WWW22}.

Despite the performance benefits, mobile cloud computing can result in privacy concerns as the input data (e.g., images) on the mobile device often involve privacy-sensitive components, which either the users themselves do not intend to release from their private possession~\cite{Zhao-arxiv23} or disallow data sharing and dissemination due to privacy regulations~\cite{GDPR,HIPAA}. Among various privacy concerns with a diverse set of data types, visual privacy~\cite{Zhao-arxiv23,Ye-MMSys22,Pad-ESA15} is one of the most direct privacy issues faced by the end users when the multimedia data (e.g., images and videos) are offloaded to the cloud for processing. Visual privacy is concerned about the visual information contained in the disseminated data being visually perceivable by unauthorized individuals other than the owner of the data, even if there are no factual security or privacy breaches or side effects that can be attributed to the visual exposure. 
Given the rich visual information in the multimedia data, the visual privacy issue has raised significant public concerns and introduced challenges in the deployments of many multimedia services~\cite{Sta-TACCESS22,Akt-Security20,Ple-arxiv23,Cha-CCWC23}.

To address the visual privacy issue, protective perturbation-based approaches have been proposed recently, which inject noises to the private images in order to prevent the visual privacy exposure to human vision while preserving the capability of machine vision to complete the machine learning services~\cite{Ren-ECCV18,Sun-CVPR18,Wu-MobiCom21,Zhu-AIES20,Ye-MMSys22}. However, such perturbation injection approaches often require executing a computation-intensive perturbation generation model on the resource-constrained mobile device, posing significant challenges to the latency, power consumption, and user experience. Therefore, it is challenging to deploy such privacy protection approaches on the mobile device in an end-to-end system. 

We develop a protective perturbation-based multimedia mobile computing framework, namely \pmc{}, to address the aforementioned privacy and resource challenges. \pmc{} adopts the state-of-the-art protective perturbation approach to protect the privacy of the private images but offloads the computation-intensive perturbation generation to a secure edge server to alleviate the computation workload on the resource-constrained mobile device. The protected images generated by the secure edge server are then transferred to the remote cloud server for the desired multimedia service. To ensure the confidentiality and integrity of the perturbation generation process, \pmc{} employs hardware-based confidential computing techniques to deploy and execute the perturbation generation model in a secure container on the edge server, which provides the same level of trustworthiness as the user mobile device and thus eliminate the potential visual privacy exposure. \tang{It is worth noting that the secure edge server is deployed in a local and user-trusted domain, while the cloud server is deployed in an untrusted, remote domain. A trusted domain is deployed with a series of security/privacy-preserving techniques, e.g., force HTTPS, access control, secure data storage and management, and firewalls to prevent security/privacy threats~\cite{TrustedNetwork}. The user trusts such a domain and has the ability to verify the integrity of the system (e.g., via attestation~\cite{SEVSNP-Attestation}). The trusted domain resides in a local area network (LAN) as small as a home LAN to deploy applications like \blue{an AI-powered photo management system~\cite{photoprism}}. In addition to the security/privacy benefits, the local trusted domain also enables faster and cheaper in-domain communications between the mobile device and the edge server. On the other hand, the untrusted domain resides in the remote cloud service, which provides powerful computation resources and proprietary machine learning models; however, it inevitably incurs bandwidth expensive cross-domain communications between the secure edge server and the cloud server.
}

While deploying the proposed \pmc{} framework, we identify a key technical challenge that the perturbed images cannot be efficiently compressed by the standard image encoding/compression algorithms, resulting in significantly increased bandwidth consumption \tang{in the cross-domain communication} when transmitting the privacy-preserving, perturbed images. To address this bandwidth challenge, we develop a neural compressor tailored to efficiently compress images with protective perturbations. The neural compressor is trained using a dataset of perturbed images and optimizes for the compression ratio while minimizing the distortion between the original and reconstructed images. Given the unique patterns of protective perturbations embedded in the images, the neural compression approach is capable of achieving a significantly higher compression ratio compared to standard image encoding methods.

We implement the proposed \pmc{} framework in an end-to-end multimedia mobile cloud computing system, involving a mobile device, a protective perturbation generator deployed in a secure edge sever with confidential containers enabled by AMD SEV~\cite{AMD-SEV}, and a regular cloud server running image recognition services. Our evaluations using 5000 images from the CIFAR-10 dataset show superior latency, power, and bandwidth results of the proposed \pmc{} framework in comparison with the baseline systems. Note that the perturbation generation and compression models required by \pmc{} can be trained with only black box access to the image recognition service model, without the knowledge of the model structure/parameters or modifying the model. Therefore, \pmc{} can be seamlessly deployed in empirical and legacy multimedia mobile cloud computing systems. 

To summarize, we have made the following contributions by developing \pmc{}:
\begin{itemize}[leftmargin=*]
\item We develop the first \black{cloud-based} protective perturbation generation system for visual privacy protection, alleviating the computation workload on the resource-constrained mobile devices to improve the latency and power efficiency;
\item The proposed neural compression approach tailored to perturbed images achieves significantly higher bandwidth savings than standard image encoding methods; and
\item We deploy the proposed \pmc{} framework in an end-to-end multimedia mobile cloud computing system, which we plan to open source upon the publication of the paper to motivate further research in the community.
\end{itemize}

\section{Background and Related Work}
\subsection{Multimedia Mobile Cloud Computing}
\label{sec:mmcc}
A typical multimedia mobile cloud computing system involves a mobile client and a cloud server. The mobile client offloads input data to the cloud server to request computation-intensive services from the cloud server, which makes the mobile device achieve significantly lower resources (e.g., power) consumption and higher performance. Also, it enables the service providers to deploy the proprietary image recognition models in their trusted domains for effective maintenance and intellectual property protection.

However, as a trade-off, the empirical deployment of such a mobile cloud computing system faces the following two challenges: 

\begin{itemize}[leftmargin=*]
\item \textbf{\emph{Privacy/Security challenge}}. Since the user data, which are often of a privacy-sensitive nature, must be offloaded to the cloud server, privacy concerns are often raised due to both the user's natural psychological needs and the legal regulations~\cite{Privacy-Blog20}. Also, the untrusted or unsecured cloud service provider and network may enable adversaries to obtain and abuse the private user data to issue various cybersecurity attacks~\cite{StrongDM-Blog23}. 

\item \textbf{\emph{Bandwidth challenge}}. The data offloading from the mobile client to the cloud server may consume a large amount of bandwidth in the communication network, especially when rich multimedia content is involved (e.g., high-definition images or videos). 
\end{itemize}

\subsection{Privacy-Preserving Protective Perturbation}
We only consider visual privacy as the threat model in this work, and do not consider security threats like crossmatching attacks or other attacks that may compromise the visual privacy solutions (e.g., protective perturbations). Many research efforts have focused on privacy-preserving image delivery to address the visual privacy challenge, where the goal is to prevent the user's private image from being exposed to others. For example, several research works propose to inject perturbations to blur and protect the privacy-sensitive information in the images~\cite{Ren-ECCV18,Sun-CVPR18,Wu-MobiCom21,Zhu-AIES20,Ye-MMSys22}. Such perturbation generator can be deployed on the mobile client to blur the sensitive images before it is exposed to the untrusted network or service provider.

Without loss of generality, we adopt the state-of-the-art protective perturbation generator~\cite{Ye-MMSys22} in our mobile cloud image recognition system. \blue{This perturbation generator utilizes U-Net $U(\cdot)$ as the core generative model, which} is trained to inject perturbations to the original images, so that they are entirely blurred to human vision for privacy protection but remain as effective inputs to the image recognition model on the cloud server to achieve highly accurate recognition results. \blue{The goal is achieved by minimizing the loss function $\omega_1 \cdot loss(y_{i_{target}}, y_i) - \omega_2 \cdot loss(y_{i_{aux}}, y_i)+\omega_3 \cdot SSIM(x_i, x_i^\prime)$, where $\omega_j\ (j \in \{1, 2, 3\})$ are tunable weights of each term, $loss(\cdot)$ is the Cross-Entropy Loss, $y_i$ is the true label of the input image $x_i$, $y_{i_{target}}$ and $y_{i_{aux}}$ are predicted labels of the perturbed image $x_i^\prime$. Note that this work utilizes two image recognition models during the training process, a target model (outputs $y_{i_{target}}$) and an auxiliary model (outputs $y_{i_{aux}}$), where the target model is the model on the cloud server, and the auxiliary model is a different image recognition model to help generate the perturbation, chosen by experiments.}

\begin{table}[htbp]
\centering
\caption{Results of using standard PNG and JPEG to compress regular and perturbed images.}
\label{tab:ineffective_compression}
\renewcommand{\arraystretch}{1.1}
\small
\begin{tabular}{cccccc}
\hline
\multirow{3}{*}{Image Type} & \multicolumn{5}{c}{Encoding Methods}                               \\ \cline{2-6} 
                            & \multicolumn{2}{c}{PNG} & \multicolumn{3}{c}{JPEG (no subsampling)}                 \\ \cmidrule(l){2-3} \cmidrule(l){4-5}
                            & Size (Bytes) & Accuracy & Size (Bytes) & Accuracy  \\ \hline
Regular             & 2662.60      & 94.24\%  & 724.44     & 90.28\%              \\
Perturbed            & 2690.56      & 91.44\%  & 1818.77      & 90.40\%             \\ \hline
\end{tabular}
\end{table}

\subsection{Limitations of Existing Protective Perturba-
tion and Image Encoding Approaches}
We conduct two case studies to reveal the limitations of the existing protective perturbation and image encoding approaches, which cannot effectively address the aforementioned privacy and bandwidth challenges in our target end-to-end mobile cloud image recognition system.

\subsubsection{Case Study 1: Excessive on-device power consumption and latency}
\begin{figure}[htbp]
\centering
\includegraphics[width=0.4\textwidth]{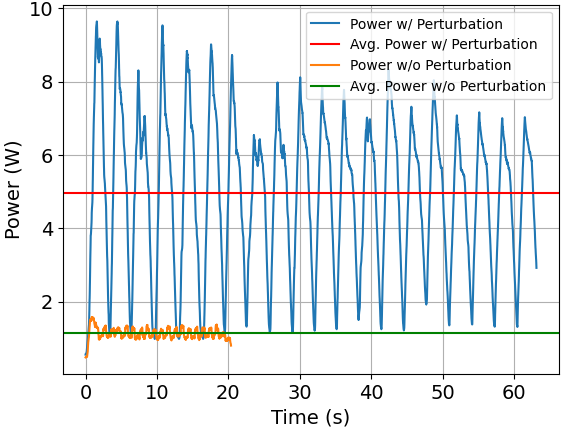}
\caption{On-device power traces with \& without privacy-preserving perturbation generation in the mobile cloud image recognition system.} 
\label{fig:power_trace}
\end{figure}

Since the perturbation generation process involves the execution of a computation-intensive neural network model, the perturbation generator is expected to consume a significant amount of battery power if deployed on the mobile device as in the state-of-the-art solutions~\cite{Ye-MMSys22}. In this case study, we measure and compare the power consumption of the image recognition application~\cite{Ye-MMSys22} running on a Pixel 3 phones \emph{with} and \emph{without} injecting protective perturbation, as shown in Figure~\ref{fig:power_trace}. The image recognition application involves a stream of 5000 test images from the CIFAR-10 dataset~\cite{cifar10}, with a batch size of 256 for the perturbation generator. The target image recognition model is VGG13\_bn, and the image encoding method is PNG. We observe that the average power consumption in the \emph{with} perturbation case \black{(the red line) is 4.961W, which is a 4.36x of 1.139W} in the \emph{without} perturbation case \black{(the green line)}. Figure~\ref{fig:power_trace} also shows the latency results of the image recognition application for 5000 test images, as indicated by the durations of the power traces. When the protective perturbation is added, the latency increases from \black{20.033s to 63.345s}, which is a \black{3.16x} slowdown. Furthermore, combining the power and latency results, the energy consumption of the privacy-preserving version of the application is around \black{13.78x} of the original application without privacy protection. To summarize, the huge overhead of protective perturbation on mobile devices (i.e., \black{4.36x in power, 3.16x in latency, and 13.78x in energy}) must be addressed to make the privacy protection approach practical.

\subsubsection{Case Study 2: Ineffective compression for perturbed image} 
\label{sec:case_study2}
The standard image encoding mechanisms like PNG and JPEG are designed for regular images optimizing for perceivable visual qualities, instead of the perturbed images targeting machine learning services in the privacy-preserving image recognition application. To identify the potential limitations, we conduct a case study comparing the effectiveness of compressing regular and perturbed images using PNG and JPEG, with the target of achieving at least 90\% of accuracy in image recognition, using as VGG13\_bn as target image recognition model and 5000 test images from the CIFAR-10 dataset~\cite{cifar10}. Table~\ref{tab:ineffective_compression} shows the results of the case study. JPEG achieves 3.7x of compression ratio on regular images compared to lossless PNG. However, such compression ratio would decrease to 1.5x in the case of perturbed images in order to maintain a similar image recognition accuracy. The significant reduction in compression ratio indicates the ineffective compression achieved by the standard image encoding methods, which would cause higher bandwidth consumption in the delivery of the perturbed images.

In summary, the two case studies reveal the limitations of existing perturbation generation and image encoding approaches in achieving a practical end-to-end privacy-preserving image recognition system, which we aim to address in this work.

\label{subsubsec:ineffective_compression}

\section{\pmc{} System Design}

We develop a privacy-preserving multimedia mobile cloud computing framework, namely \pmc{}, to proactively protect the privacy of user data in mobile-cloud image recognition, while consuming low on-device resources and network bandwidth for practical deployment. In a nutshell, the \pmc{} system addresses the aforementioned \textbf{\emph{privacy/security challenge}}, as well as the associated power/latency overhead, by executing the security-sensitive and computation-intensive perturbation generation task on the trusted \emph{secure edge server}, which is protected by the state-of-the-art confidential computing techniques and equipped with abundant resources for high performance computing. Also, it addresses the aforementioned \textbf{\emph{bandwidth challenge}} by employing perturbation-aware neural compression for the perturbed images. Furthermore, the \emph{deployment} of \pmc{} system does not require modifications to the target image recognition model, which is assumed to be proprietary, making it immediately deployable to empirical mobile cloud systems.

\subsection{System Overview}
\begin{figure}[htbp]
\centering
\includegraphics[width=0.48\textwidth]{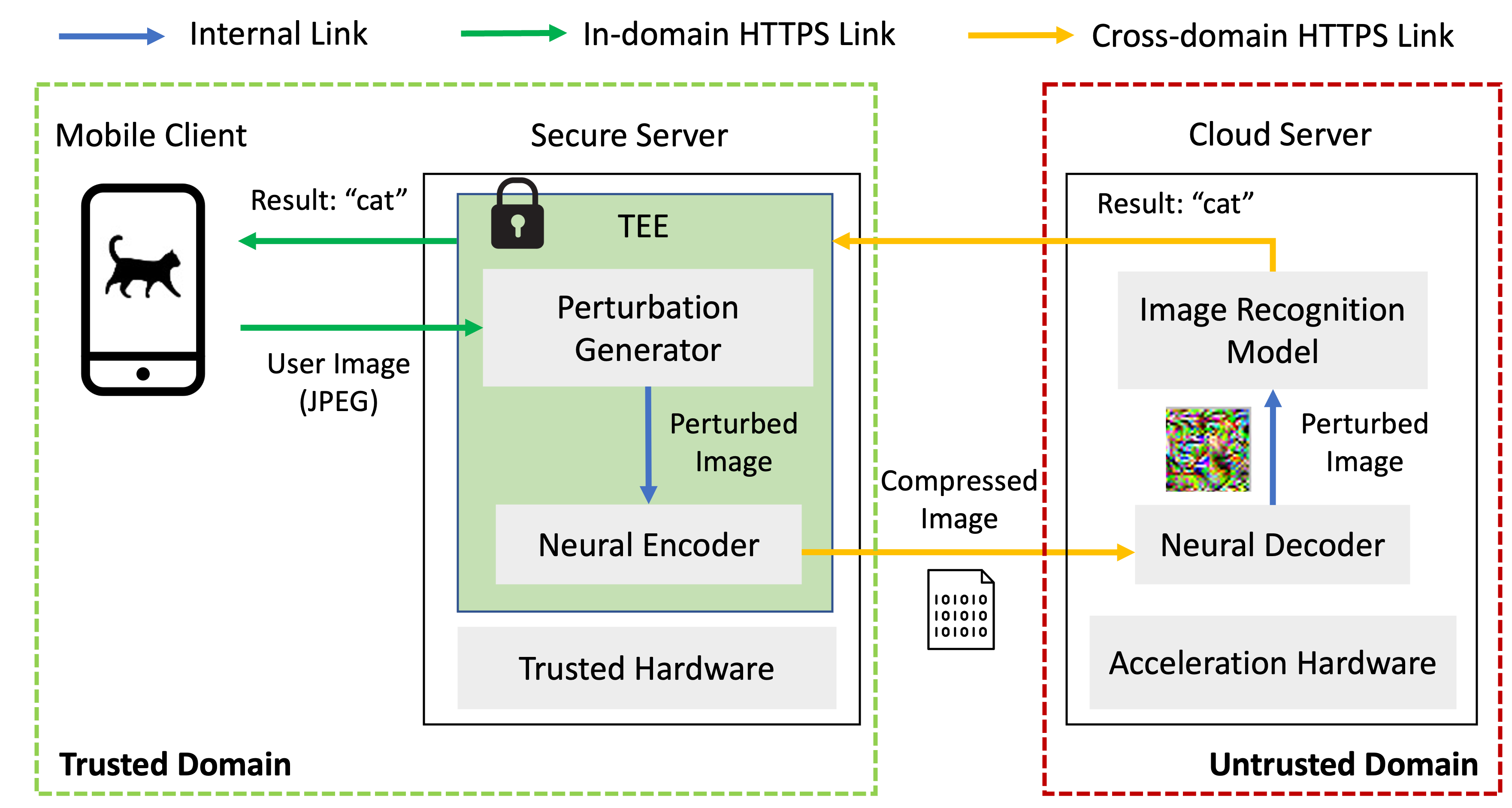}
\caption{End-to-end workflow of the proposed \pmc{} system.} 
\label{fig:sys}
\end{figure}

\black{Figure~\ref{fig:sys} shows the overall architecture of \pmc{}. It introduces a trusted \emph{secure edge server} between the \emph{mobile client} and the traditional \emph{cloud server}. The \emph{secure edge server} accomplishes privacy-preserving perturbation generation in a confidential computing container protected by the trusted execution environment (TEE) technology. In addition, a neural encoder and decoder are deployed on the \emph{secure edge server} and the \emph{cloud server}, respectively, to attain the bandwidth-saving goal.}

The end-to-end \pmc{} system works as follows to achieve privacy-preserving mobile-cloud image recognition. \emph{First}, the \emph{mobile client} \black{compresses the privacy-sensitive images using JPEG and offloads them} to the trusted \emph{secure edge server} \tang{via an in-domain HTTPS link. The \emph{secure edge server} is deployed inside the same trusted domain as the mobile client}. \emph{Second}, on the \emph{secure edge server}, a perturbation generator runs inside the TEE-protected confidential container to securely inject protective perturbations to the received user images in real time; Then, the perturbed images are compressed by a perturbation-aware \emph{neural encoder} and transferred to the \emph{cloud server} for image recognition \tang{via the cross-domain HTTPS link}. \tang{The \emph{cloud server} is deployed in a remote and untrusted domain.} \emph{Third}, the \emph{cloud server} decodes the compressed images using a \emph{neural decoder}, which reconstructs the perturbed images and feeds them into the machine learning-based image recognition model for the computation-intensive classifications. \emph{Finally}, the image recognition results are returned to the \emph{mobile client} via the \emph{secure edge server}.

\subsection{Secure Protective Perturbation Generation}
\label{subsec:secure_pertgen}

We employ the state-of-the-art confidential computing techniques to generate the protective perturbations on the \emph{secure edge server} to address the \textbf{\emph{privacy/security challenge}} without incurring power and latency overhead on the \emph{mobile client}. Confidential computing has been widely adopted in the modern cloud platforms by leveraging trusted computing environment (TEE) technology ~\cite{AWS-CC,Google-CC,Azure-CC,IBM-CC}. Such technology provides secure enclaves in the system, which guarantees that the computation conducted in each enclave is encrypted and isolated from the rest of the system. Generally there are two types of TEEs, namely process-based TEEs and virtual machine (VM)-based TEEs. Process-based TEEs such as Intel SGX~\cite{sgxref} can achieve minimal trusted computing base (TCB) since they only deploy the processes that must be protected into enclaves. On the other hand, such techniques require partitioning each application into secure and non-secure domains and thus introduce significant offline overhead (i.e., application partitioning) and runtime overhead (i.e., context switches between the secure and non-secure domains). Due to these limitations on process-based TEEs, there has recently been increasing demand for VM-based TEEs, such as Intel TDX~\cite{Intel-TDX} and AMD SEV~\cite{AMD-SEV}, which deploy the entire VM into an enclave. Although VM-based TEEs result in larger TCB than the process-based TEEs, they are more suitable for cloud-native computing and can be easily integrated into Infrastructure as a Service (IaaS) and Platform as a Service (PaaS) platforms.

To protect the confidentiality of the perturbation generation in \pmc{}, we deploy the perturbation generator into a TEE on the \emph{secure edge server}. Note that our proposed framework is generally compatible with any TEE techniques but, in this work, we leverage confidential containers (CoCo)~\cite{CoCo} with AMD SEV~\cite{AMD-SEV} (i.e., a VM-based TEE) for the following reasons. \emph{First}, CoCo supports running containerized applications and can be managed by Kubernetes (K8s)~\cite{K8s} platforms, which helps us achieve a cloud-native computing environment. \emph{Second}, CoCo supports multiple VM-based TEE techniques, such as AMD SEV and Intel TDX~\cite{Intel-TDX}, which provides flexibility for the users to choose different hardware platforms. \emph{Third}, applications can be directly deployed into CoCo without any modifications, which significantly increases the diversity of supported perturbation generation models that can be applied to the \pmc{} framework and does not introduce additional offline overhead to modify the models.

Kindly notice that the protective perturbation generators run with CPU only without leveraging GPU or other acceleration hardware since accelerator TEEs have not become widely available in the server/cloud environment yet. We will leverage secure accelerators in the future upon their availability.

\subsection{Bandwidth-Saving Encoding}

As shown in Figure~\ref{fig:sys}, there are two communication links in \pmc{} that consume network bandwidth: 
(a) \tang{in-domain,} the compressed user image is transmitted from the \emph{mobile client} to the \emph{secure edge server}, and (b) \tang{cross-domain,} the encoded perturbed image is transmitted from the \emph{secure edge server} to the \emph{cloud server}. Let us denote the original image as $Img_{orig}$, which is encoded and stored in PNG by default. The JPEG-compressed version, $JPEG(Img_{orig})$, is delivered via \tang{the in-domain link}. The perturbed image $Img_{pert}$ is equivalent to $PertGen(JPEG(Img_{orig}))$, where $PertGen(\cdot)$ is the perturbation generator, and it will be encoded by the encoder function $Enc(\cdot)$. Then, $Enc(Img_{pert})$ is the image transmitted via \tang{the cross-domain link}.

\black{Based on our experiments, the target model accuracy for \\ $PertGen(JPEG(Img_{orig}))$ and $PertGen(Img_{orig})$ have very small differences (< 2\%) with huge bandwidth savings (about 40\%); therefore, the bandwidth-saving encoding for the \tang{the in-domain link} can be easily solved by adopting JPEG.} To address the \emph{bandwidth challenge} for \tang{the cross-domain link}, i.e., \black{$Enc(Img_{pert})$, 
} 
we consider PNG and JPEG as the baseline $Enc(\cdot)$ approaches, which are subject to the compression inefficiency issue discussed in Section~\ref{sec:case_study2}.  PNG achieves lossless compression but with low compression ratio and high bandwidth consumption. For JPEG, we observe that the default $subsampling$ parameters would break the functionality of the target model. When $subsampling$ is disabled, to attain the same level of the target model accuracy, the compression ratio for perturbed images is significantly lower than that of regular images, as shown in Section~\ref{sec:case_study2}. These observations motivate us to develop a new neural compressor tailored to the perturbed images to reduce the bandwidth consumption on \tang{the cross-domain link}. 

\subsubsection{Baseline Approach: JPEG Compression}

\black{4:2:0 chroma subsampling is widely used in video cameras, Blu-ray movies, and also used by default in libjpeg-turbo~\cite{libjpeg-turbo}, a popular JPEG library. The first number 4 specifies a 4x2 pixel block, the second number 2 means there are two chroma samples in the first row, and the last number 0 stands for the second row of chroma values being discarded. This subsampling makes chroma samples one-quarter of the luma ones. In our case study that uses JPEG to compress the protected image generated by the protective perturbation, we observe the target model accuracy reduces signifiantly with chroma subsampling enabled. For example, when the quality factor is 89, the target model accuracy decreases from 91.17\% to 11.50\%, and the file size decreases from 2057.36 B to 1212.33 B. Although the file size is \black{significantly reduced}, the functionality of the target model is compromised. This indicates that the image recognition model recognizes the luma and chroma channels of the protected images equally and, when a lot of information is discarded in the chroma channel, the model cannot work properly. To maintain the high accuracy of the image recognition model and reduce the file size simultaneously, we disable the chroma subsampling (i.e., the sampling factor is 4:4:4) when using the JPEG encoding. The quality factor used in the system can be found in Table~\ref{tab:accuracy_bandwidth}. The PNG and JPEG compression/decompression are condcuted by the Python Pillow 8.4.0~\cite{pillow}, with the $optimize$ option set as $True$.}

\subsubsection{Proposed Approach: Perturbation-Aware Neural Compression}

\black{Neural image compression~\cite{Yang-FTCGV23} is an emerging approach to image compression that employs neural networks to compress and decompress images. Unlike traditional image compression methods, which typically rely on heuristics and mathematical algorithms, neural image compression uses machine learning techniques to learn a compressed representation of the image. A typical lossy neural compressor works as follows.
The input image $x$ is first mapped to a latent representation $y$ 
by an encoding transformation $f$ (i.e., a neural network).
In the next step, $y$ is quantized to $\hat{y}$, and then a lossless entropy encoding model (e.g., EntropyBottleneck~\cite{Balle-arXiv18})
encodes it to a bit-string. During decompression, the aforementioned steps are executed in the reverse order to generate the reconstructed image $\tilde{x}$. Since the compression is lossy, an error between $x$ and $\tilde{x}$ will be introduced.}

\begin{figure}[htbp]
\centering
\includegraphics[width=0.45\textwidth]{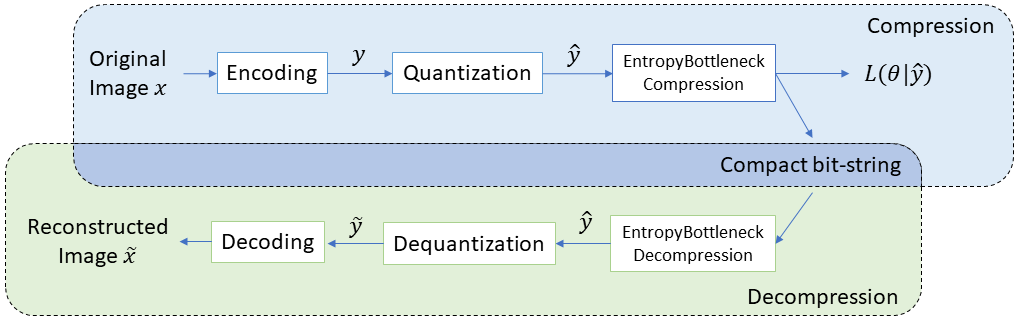}
\caption{Workflow of the perturbation-aware neural compression and decompression.}
\label{fig:nc_arch}
\end{figure}

\black{Figure~\ref{fig:nc_arch} illustrates the workflow of the neural compression and decompression procedure in \pmc{}, which is based on the Factorized Prior model~\cite{Balle-arXiv18}. In compression, the input image $x$ is first encoded to latent $y$. The encoding step consists of three convolution layers, each followed by a Generalized Divisive Normalization (GDN) layer~\cite{Balle-arXiv15}. $y$ is quantized to $\hat{y}$ to reduce the entropy (or the bit-rate). The EntropyBottleneck (EB) layer~\cite{Balle-arXiv18} utilizes the built-in probabilistic model to generate the compressed bit-string and the likelihood of $\hat{y}$ $L(\theta|\hat{y})$ under the parameters $\theta$ of the EB layer. In decompression, the compressed bit-string is decompressed by the EB first to recover $\hat{y}$, and the quantization and the decoding steps convert it to the reconstructed image $\tilde{x}$. The decoding step is the inverse of the encoding, and it contains three deconvolution layers, each of which is followed by an inversed GDN layer.}

\black{In theory, the neural compressor used for \pmc{} (i.e., high compression ratio, low impact on the target model accuracy) should be trained by a loss function comprised of rate loss and the accuracy loss. \black{In this case, there is no need to optimize for distortion loss as it indicates the perceivable visual quality of a compressed image, which is not a concern in the context of compression for a machine learning service. The design of protective perturbation has resulted in a poor visual quality (most SSIM values between the original and the perturbed images are less than 0.02)~\cite{Ye-MMSys22}, which is desired and unlikely to 
be restored and break the visual privacy protection by the neural compressor.} In the experiments, we find that using the classical rate-distortion loss without any modification for training gives us a compressor that meets our \black{rate-accuracy} goal. In the rate-distortion loss, the loss comes from the mean square error (MSE) between $x$ and $\tilde{x}$, and we use \black{bits per pixel (bpp)} as the rate loss calculated by}

\begin{equation}
    rate\_loss = - \frac{\sum(\log_2 L(\theta|\hat{y}))}{num\_pixels}
\end{equation}

\black{\noindent where $num\_pixels$ is the total number of pixels in $x$. To train the Factorized Prior model using the CIFAR-10 dataset~\cite{cifar10} perturbed by the protective perturbation generator, we set the rate-distortion parameter $\lambda$ as 0.01. Adam optimizer~\cite{Kingma-arXiv14} is used to optimize both the rate-distortion loss of the model and the auxiliary loss in the Entropy Bottleneck layer, with 0.0001 and 0.001 learning rates, respectively. ReduceLROnPlateau~\cite{ReduceLROnPlateau} is used as the lr scheduler for the rate-distortion loss, and works in the min mode, which means that the lr will be reduced when the loss stops decreasing. Moreover, the gradient norm will be clipped when exceeding 1.0 during the training. The training batch size is set as 16, the number of workers (dataloader threads) is set as 4, and the maximum training epochs is 100. In the evaluation, the batch size is 256. The neural compressor is trained and deployed with the help of the Compress AI framework~\cite{Begaint-arXiv20}.}

\section{Experimental Results}

\subsection{Experimental Setup and Implementation}
\label{sec:experimental_setup}

\subsubsection{\pmc{} System Setup}
\label{sec:pmc_system_setup}
We deploy the end-to-end system as illustrated in Figure~\ref{fig:sys} to evaluate the performance of the proposed \pmc{} system. The \emph{mobile client} is a Google Pixel 3 smartphone running Android 12, which has four 2.5GHz CPUs, four 1.6GHz CPUs, an Adreno 630 GPU, and a 2915 mAh battery. A client APP runs on the mobile device to send the \black{compressed JPEG images (by Android \emph{Bitmap.compress()} API)} and receive the recognition results. 
The \emph{secure edge server} has an AMD EPYC 7443P Milan CPU with 24 cores and 48 threads, runs at 2.9GHz, and equips with 256 GB memory. Following the implementation of \emph{secure edge server} described in Section~\ref{subsec:secure_pertgen}, we adopt and deploy the opensource \emph{perturbation generator}~\cite{Ye-MMSys22} to the confidential container of the \emph{secure edge server}. We choose 8 target image recognition models and train their corresponding perturbation generators following the instructions provided in~\cite{Ye-MMSys22}. The auxiliary model used in the training and the target model accuracy of the test images are listed in Table~\ref{tab:generator}. We run all the evaluations of generators using the batch size of 256. The target image recognition model runs on a \emph{cloud server} with \black{8 vCPUs, 1/3 NVIDIA A40 GPU, and 40 GB memory}. 
The data flows between the client and the two servers are transmitted over HTTPS.
\black{The \emph{mobile client} is deployed locally  \blue{and connects to the Internet via an 802.11ac Wi-Fi}, and the \emph{secure edge server} and the \emph{cloud server} are deployed in different cloud centers~\cite{Vultr}. Specifically, the \emph{secure edge server} is located much closer to the \emph{mobile client} than the \emph{cloud server} to serve as an edge server.
}

\subsubsection{Baseline System Setup} We adopt the opensource protective perturbation system developed in \cite{Ye-MMSys22} as the baseline system for comparison. For fair evaluation and comparison, we upgrade the communication protocol in the baseline system to \black{HTTPS} and change the generator's batch size from 1 to 256. The mobile client APP and the server program run on the same mobile device and the cloud server as the \pmc{} system.

\begin{table}[htbp]
\centering
\caption{The target models and auxiliary models used to train the perturbation generators, and the target model accuracy when applying the corresponding perturbation generator to the original test images.}
\label{tab:generator}
\renewcommand{\arraystretch}{1.1}
\small
\begin{tabular}{ccc}
\hline
Target   Model & Auxiliary Model & \begin{tabular}[c]{@{}c@{}}Target Model\\ Accuracy\end{tabular} \\ \hline
VGG13\_bn      & ResNet18        & 91.52\%                                                          \\
VGG16\_bn      & ResNet18        & 91.94\%                                                         \\
ResNet18       & VGG13\_bn       & 91.62\%                                                          \\
ResNet34       & ResNet18        & 91.54\%                                                          \\
DenseNet121    & ResNet18        & 91.38\%                                                          \\
MobileNet\_v2  & VGG13\_bn       & 91.82\%                                                          \\
GoogLeNet      & ResNet18        & 91.22\%                                                          \\
Inception\_v3  & GoogLeNet       & 91.86\%                                                          \\ \hline
\end{tabular}
\end{table}

\subsubsection{Dataset} We use the CIFAR-10~\cite{cifar10} dataset to train all the models and perform all the evaluations. \blue{The images are in the size of 32x32 and the average file size is 2662.56 Bytes.} We use the 50000 images in the training set of CIFAR-10 to train the 8 protective perturbation generators \black{listed in Table~\ref{tab:generator}}, 5000 images in the test set for validation, and the rest 5000 images in the test set for evaluations. We split the test dataset into two 5000-image sets in the same way as described in \cite{Ye-MMSys22}. \black{For each perturbation generator, we use the same images but with the particular perturbations added by the generator to train a dedicated Factorized Prior neural compression model~\cite{Balle-arXiv18}.}

\begin{table*}[htbp]
\centering
\renewcommand{\arraystretch}{1.1}
\small
\caption{Accuracy and bandwidth comparison for different encoding methods across all target models. \black{$S_1$ and $S_2$ represent \tang{the average file sizes over all 5000 images} from the \emph{mobile client} to the \emph{secure edge server} and from the \emph{secure edge server} to the \emph{cloud server}, respectively.}}
\label{tab:accuracy_bandwidth}
\begin{tabular}{cccccccccc}
\hline
\multirow{3}{*}{Target Model} & \multicolumn{9}{c}{Encoding Method}                                                                          \\ \cline{2-10} 
                              & \multicolumn{3}{c}{PNG}            & \multicolumn{3}{c}{JPEG}           & \multicolumn{3}{c}{Neural}         \\ \cmidrule(l){2-4} \cmidrule(l){5-7} \cmidrule(l){8-10} 
                              & $S_1$ (Bytes) & $S_2$ (Bytes) & Accuracy & $S_1$ (Bytes) & $S_2$ (Bytes) & Accuracy & $S_1$ (Bytes) & $S_2$ (Bytes) & Accuracy \\ \hline
VGG13\_bn   & 1642.45$\pm$6.42 & 2691.38$\pm$6.79& 90.30\%  & 1854.95$\pm$6.69& 1848.42$\pm$1.12& 90.22\%  & 1642.45$\pm$6.42& 8.16$\pm$0.05& 90.40\%  \\
VGG16\_bn   & 1509.06$\pm$5.79 & 2599.11$\pm$3.13& 90.06\%  & 1572.23$\pm$6.17& 1984.75$\pm$0.41& 90.14\%  & 1509.06$\pm$5.79& 8.01$\pm$0.02& 90.06\%  \\
ResNet18    & 1509.06$\pm$5.79 & 2544.70$\pm$3.27& 90.22\%  & 1509.06$\pm$5.79& 1861.48$\pm$0.35& 90.24\%  & 1509.06$\pm$5.79& 8.12$\pm$0.03& 90.20\%  \\
ResNet34    & 1572.23$\pm$6.17 & 2576.23$\pm$3.36& 90.04\%  & 1726.20$\pm$6.73& 1943.05$\pm$0.48& 90.12\%  & 1642.45$\pm$6.42& 8.19$\pm$0.05& 90.42\%  \\
DenseNet121   & 1572.23$\pm$6.17 & 2794.40$\pm$2.14& 90.48\%  & 1642.45$\pm$6.42& 2042.20$\pm$0.65& 90.38\%  & 1572.23$\pm$6.17& 8.29$\pm$0.07& 90.42\%  \\
MobileNet\_v2  & 1509.06$\pm$5.79 & 2636.92$\pm$7.65& 90.20\%  & 1572.23$\pm$6.17& 2113.56$\pm$1.51& 90.26\%  & 1572.23$\pm$6.17& 8.35$\pm$0.08& 90.58\%  \\
GoogLeNet      & 1572.23$\pm$6.17 & 2796.73$\pm$5.53& 90.00\%  & 1726.20$\pm$6.73& 1891.50$\pm$1.06& 90.32\%  & 1726.20$\pm$6.73& 8.00$\pm$0.01& 90.42\%  \\
Inception\_v3  & 1509.06$\pm$5.79 & 3170.88$\pm$0.11& 90.32\%  & 1509.06$\pm$5.79& 2390.50$\pm$0.57& 90.04\%  & 1572.23$\pm$6.17& 8.00$\pm$0.00& 90.36\%  \\ \hline
\end{tabular}
\end{table*}

\subsection{Accuracy and Bandwidth Overhead}

\black{Table~\ref{tab:accuracy_bandwidth} presents the image recognition accuracy and the average size of the compressed image when deploying different image encoding methods between the \emph{secure edge server} and the \emph{cloud server}. The average size of $JPEG(Img_{orig})$ (i.e., the images sent from the \emph{mobile client} to the \emph{secure edge server}) and the average size of $Enc(Img_{pert})$ (i.e., the images sent from the \emph{secure edge server} to the \emph{cloud server}) are represented by $S_1$ and $S_2$, respectively. \tang{The standard deviation values of $S_1$ and $S_2$ are also provided after the $\pm$ symbols to describe the distribution of file sizes. All the standard deviation values are relatively small, indicating all the file sizes are generally close to the average values.}
The PNG-encoded $Img_{pert}$ achieves recognition accuracies between 90.00\% to 90.48\%, and the image sizes range from 2544.70 to 3170.88 Bytes. The accuracy values decrease by 0.90\% to 1.88\% compared to the numbers reported in Table~\ref{tab:generator}. Since the results in Table~\ref{tab:generator} come from the cases where uncompressed user images are used to generate perturbed images, $JPEG(Img_{orig})$ (i.e., the reduction of $S_1$) is the only factor that impacts the target model accuracy. Considering the raw size of a 32x32 RGB image is just 3096 Bytes (32x32x3), it is apparent that the protected images are not effectively compressed in the PNG format.} \black{We then take a step forward by replacing PNG with JPEG compression, which achieves significant data size reduction, ranging from 19.85\% (MobileNet\_v2) to 32.37\% (GoogLeNet), while maintaining similar recognition accuracies (i.e., above 90\%). }

\black{The neural compressor in the proposed \pmc{} system shows superior accuracy and bandwidth savings. More specifically, the neural-compressed images still keep high target model accuracy, which is only reduced by 0.80\% (GoogLeNet) to 1.88\% (VGG16\_bn) compared to the PNG encoding. On the other hand, a protected image can be compressed to a bit-string ranging from 8 to 8.35 Bytes, which is achieved by a neural compression model sized at only 12 MB. The data size is reduced by more than 99\% compared to both PNG and JPEG encoding methods. In short, the proposed neural compression outperforms JPEG compression in terms of  data size and thus bandwidth consumption between the \emph{secure edge server} and the \emph{cloud server}, without affecting the functionality of the image recognition model. }

\black{Note that the quality factors used for JPEG in both links, if applicable, are chosen experimentally to achieve similar target model accuracy (i.e., 90\%) for a fair comparison.  It is also worth noting that the average size of $Img_{orig}$ is 2662.56 Bytes and, after applying JPEG compression with proper JPEG quality factors, the size (i.e., $S_1$) decreases by 40\% on average, with almost no negative impact on the target model accuracy. }

\begin{figure*}[htbp]
\centering
\includegraphics[width=0.97\linewidth]{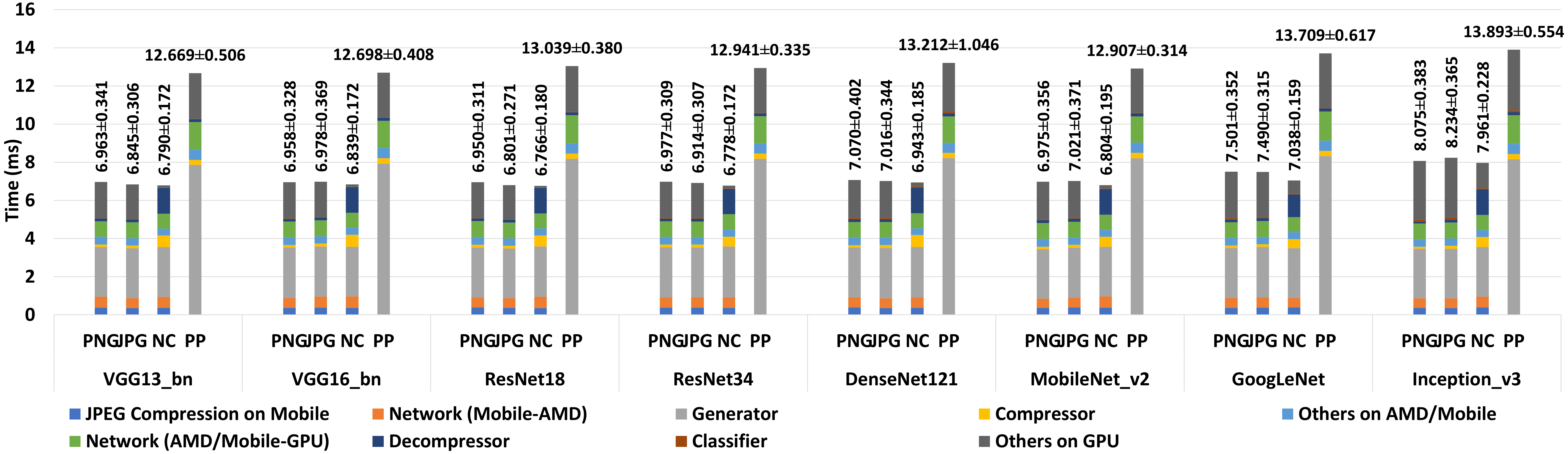}
\caption{Timing performance of the \pmc{} system and the baseline protective perturbation system, employing 8 target models. Each target model is deployed in four different ways, including PNG, JPG (JPEG), NC (Neural Compressor), and PP (baseline protective perturbation system using PNG). } 
\label{fig:timing}
\end{figure*}

\subsection{Timing Performance}
We further evaluate the end-to-end latency of the proposed \pmc{} system using the 8 target image recognition models and image encoding methods. For comparison, we also conduct the same timing evaluation on the baseline protective perturbation system to uncover the gains of the proposed mechanism. Figure~\ref{fig:timing} shows \blue{the average end-to-end latency results with standard deviation values and} timing breakdown analysis in 4 test cases for each target model. The first 3 test cases, namely \emph{PNG}, \emph{JPG}, and \emph{NC}, represent using PNG, JPEG, and neural compression for the \emph{secure edge server}-\emph{cloud server} link in the \pmc{} system; the 4th test case, namely protective perturbation (\emph{PP}), represents the baseline protective perturbation system with PNG encoding. 
\blue{The time breakdown results include nine components, which are listed at the bottom of Figure~\ref{fig:timing}. The \emph{secure edge server} is named ``AMD", and the \emph{cloud server} is named ``GPU". Particularly, there are no \jpgOnMobileName and ``Network (Mobile-AMD)" components in the baseline case (i.e., \emph{PP}).}

\black{The end-to-end latency of the \pmc{} system ranges from 6.766 to 8.234 $ms$, which is significantly less than the baseline protective perturbation system ranging from 12.669 $ms$ to 13.893 $ms$. Among all three cases for \pmc{}, the neural compression approach consistently outperforms the standard PNG by 1.40\% (Inception\_v3) to 6.17\% (GoogLeNet), and it outperforms JPEG by 0.90\% (ResNet18) to 6.42\% (GoogLeNet). In general, \pmc{} takes 42.69\% (Inception\_v3 w/ neural compressor) to 48.67\% (GoogLeNet w/ neural compressor) less time than the baseline protective perturbation system for the same target model.}

\begin{figure*}[htbp]
\centering
\includegraphics[width=0.9\linewidth]{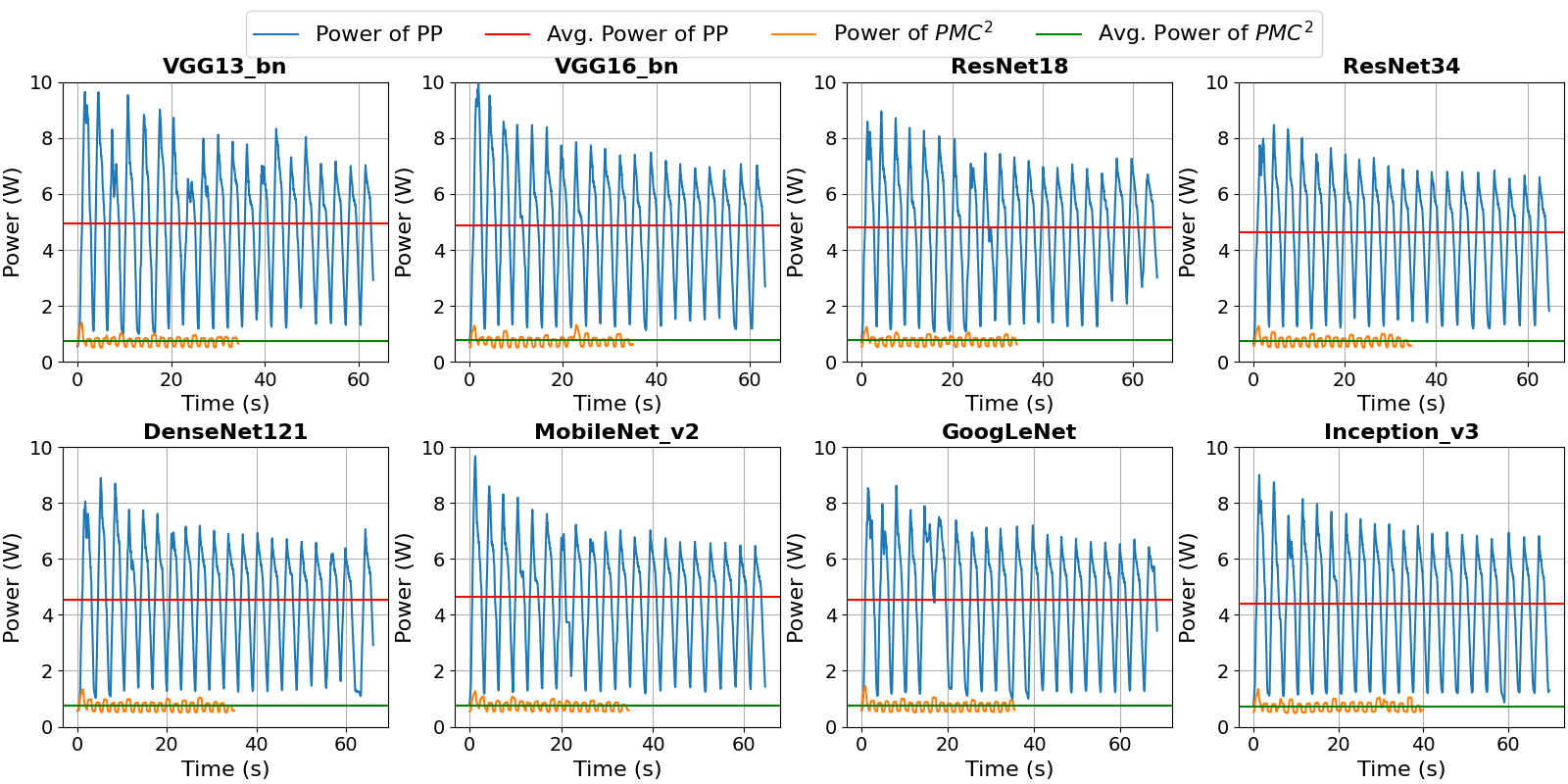}
\caption{The power performance comparison between the baseline protective perturbation (PP) system and our \pmc{} system using different target models. PNG is used in the baseline system, and neural compressor is used in \pmc{}.}
\label{fig:power}
\end{figure*}

\black{The \jpgOnMobileName\ takes about 0.37 $ms$ on average, while the ``Network (Mobile-AMD)" takes about 0.54 $ms$ on average. Based on its design, the protective perturbation generator has an almost constant running time regardless of the target image recognition model. On average, it takes 2.615 $ms$ inside an SEV-enabled CoCo container or 8.135 $ms$ on a smartphone to generate a protected image. Overall, it is the most time-consuming component in all the cases, which can take up to 39.27\% of the processing time spent in \pmc{}, and 63.60\% of that in the baseline protective perturbation system.}

\black{The average PNG compression and decompression times are 0.125 $ms$ and 0.142 $ms$ in \pmc{}, 0.272 $ms$ and 0.164 $ms$ in the baseline system, while those for JPEG are 0.150 $ms$ and 0.158 $ms$ in \pmc{}. For both PNG and JPEG, compression and decompression take less than 4.45\% of the total running time in \pmc{}, and at most 3.49\% in the baseline system. Other operations on the \emph{secure edge server} take about 0.4 $ms$ for both encoding methods, and other operations on the \emph{cloud server} take about 1.8 to 3.1 $ms$. On the other hand, the compression and decompression times are 0.558 $ms$ and \black{1.325 $ms$ 
} on average for the neural compressor, which take 7.99\% to \black{18.96\% time} in the entire system.
The neural compressor needs 0.362 $ms$ for other operations on the \emph{secure edge server} and 0.099 to 1.322 $ms$ on the \emph{cloud server}. 
It turns out that PNG and JPEG encoding/decoding take approximately equal time, with PNG slightly faster. The neural compressor needs about 4x time for compression and 9x time for decompression but less time for other operations to prepare the data for compression/decompression.} \blue{The network latency and the image recognition model take stable and short time in all cases. }

\subsection{Power Consumption}

We use the Monsoon power monitor~\cite{Monsoon} to collect the power traces of the mobile device. The power monitor is connected to a Ubuntu workstation and controlled by Python scripts. The smartphone keeps the same network connection and screen brightness (e.g., the lowest) during the testing for a fair comparison.

\black{Figure~\ref{fig:power} shows the power consumption comparison between the baseline protective perturbation system and the \pmc{} system using 8 different target models. PNG is employed for the baseline system, and the neural compressor is adopted for the \pmc{} system. The results for all the target models show a similar pattern that indicates significant power savings on the resource-constrained mobile device achieved by the proposed \pmc{} system. For example, we can take a closer look at the two power tracing results of VGG13\_bn and Densenet121 as, in the timing results of \pmc{}, VGG13\_bn is the fastest target model among all 8 models, and DenseNet121 is the slowest.} \black{In the VGG13\_bn case, 
the target image recognition model used in both systems is VGG13\_bn. In the baseline system, most power traces fall into the 1.5 to 8.0 Watts range (blue lines), averaging 4.961 Watts (red lines). 
As a comparison, the \pmc{} system consumes 0.751 Watts on average (green lines). Offloading the protective perturbation generation to the \emph{secure edge server} makes the system much faster, as the test only lasts 33.956 seconds, almost half of the 63.345 seconds for the baseline system. Suppose we calculate the energy consumption by multiplying the average power and processing time; it turns out that the \pmc{} costs only 8.11\% energy of the baseline system. Similarly, when DenseNet121 is deployed as the target model, the baseline system consumes 66.065 seconds and 4.556 Watts average power to process all the 5000 images from the test dataset. Meanwhile, \pmc{} only needs 34.716 seconds and 0.748 Watts average power to finish the same task, which yields a 91.37\% energy reduction. }

\section{Conclusion}
We have developed a privacy-preserving multimedia mobile cloud computing framework, namely \pmc{}, to address the privacy/security and bandwidth challenges in the mobile image recognition application. \pmc{} offloads the computation-intensive perturbation generation operation intended for privacy protection from the resource-constrained mobile device to the trustworthy confidential container operated on a resource-rich secure server, which achieves superior power efficiency and latency compared to the state-of-the-art mobile-only system, while maintaining the privacy and security of user images. Furthermore, \pmc{} employs a neural compression mechanism to effectively compress the perturbed images, which achieves significant bandwidth savings compared to the standard PNG and JPEG image encoding techniques. We implemented \pmc{} in an end-to-end mobile cloud image recognition system and evaluated its accuracy, bandwidth, timing, and power consumption in comparison to the baseline, mobile-only protective perturbation system. Our evaluation results based on 5000 test images and 8 target image recognition models justify the superiority of the proposed \pmc{} system. Upon the publication of this paper, we plan to open source \pmc{} on GitHub to motivate further research.

\newpage

\bibliographystyle{ACM-Reference-Format}
\bibliography{pmc2}


\begin{thebibliography}{46}


\ifx \showCODEN    \undefined \def \showCODEN     #1{\unskip}     \fi
\ifx \showDOI      \undefined \def \showDOI       #1{#1}\fi
\ifx \showISBNx    \undefined \def \showISBNx     #1{\unskip}     \fi
\ifx \showISBNxiii \undefined \def \showISBNxiii  #1{\unskip}     \fi
\ifx \showISSN     \undefined \def \showISSN      #1{\unskip}     \fi
\ifx \showLCCN     \undefined \def \showLCCN      #1{\unskip}     \fi
\ifx \shownote     \undefined \def \shownote      #1{#1}          \fi
\ifx \showarticletitle \undefined \def \showarticletitle #1{#1}   \fi
\ifx \showURL      \undefined \def \showURL       {\relax}        \fi
\providecommand\bibfield[2]{#2}
\providecommand\bibinfo[2]{#2}
\providecommand\natexlab[1]{#1}
\providecommand\showeprint[2][]{arXiv:#2}

\bibitem[GDP(2016)]%
        {GDPR}
 \bibinfo{year}{2016}\natexlab{}.
\newblock \showarticletitle{{GDPR}. Intersof Consulting}.
\newblock
\newblock
\shownote{\url{https://gdpr-info.eu}}.


\bibitem[Str(2020)]%
        {StrongDM-Blog23}
 \bibinfo{year}{2020}\natexlab{}.
\newblock \showarticletitle{40+ Alarming Cloud Security Statistics for 2023}.
\newblock
\newblock
\shownote{\url{https://www.strongdm.com/blog/cloud-security-statistics}}.


\bibitem[Pri(2020)]%
        {Privacy-Blog20}
 \bibinfo{year}{2020}\natexlab{}.
\newblock \showarticletitle{The Importance of Privacy—Both Psychological and Legal}.
\newblock
\newblock
\shownote{\url{https://www.psychologytoday.com/us/blog/emotional-nourishment/202007/the-importance-privacy-both-psychological-and-legal}}.


\bibitem[HIP(2021)]%
        {HIPAA}
 \bibinfo{year}{2021}\natexlab{}.
\newblock \showarticletitle{{HIPAA}. US Department of Health and Human Services}.
\newblock
\newblock
\shownote{\url{https://www.hhs.gov/hipaa/index.html}}.


\bibitem[sgx(2021)]%
        {sgxref}
 \bibinfo{year}{2021}\natexlab{}.
\newblock \bibinfo{title}{Intel Software Guard Extensions ({SGX})}.
\newblock
\newblock
\newblock
\shownote{\url{https://software.intel.com/content/www/us/en/develop/topics/software-guard-extensions.html}}.


\bibitem[AMD(2023)]%
        {AMD-SEV}
 \bibinfo{year}{2023}\natexlab{}.
\newblock \showarticletitle{{AMD} Secure Encrypted Virtualization}. In \bibinfo{booktitle}{\emph{AMD}}.
\newblock
\newblock
\shownote{\url{https://www.amd.com/en/processors/amd-secure-encrypted-virtualization}}.


\bibitem[SEV(2023)]%
        {SEVSNP-Attestation}
 \bibinfo{year}{2023}\natexlab{}.
\newblock \bibinfo{title}{Attestation with AMD SEV-SNP}.
\newblock
\newblock
\newblock
\shownote{\url{https://docs.aws.amazon.com/AWSEC2/latest/UserGuide/snp-attestation.html}}.


\bibitem[Azu(2023)]%
        {Azure-CC}
 \bibinfo{year}{2023}\natexlab{}.
\newblock \showarticletitle{Azure confidential computing}. In \bibinfo{booktitle}{\emph{Microsoft}}.
\newblock
\newblock
\shownote{\url{https://azure.microsoft.com/en-us/solutions/confidential-compute/}}.


\bibitem[Goo(2023)]%
        {Google-CC}
 \bibinfo{year}{2023}\natexlab{}.
\newblock \showarticletitle{Confidential Computing - {Google Cloud}}. In \bibinfo{booktitle}{\emph{Google}}.
\newblock
\newblock
\shownote{\url{https://cloud.google.com/confidential-computing}}.


\bibitem[AWS(2023)]%
        {AWS-CC}
 \bibinfo{year}{2023}\natexlab{}.
\newblock \showarticletitle{Confidential computing: an {AWS} perspective}. In \bibinfo{booktitle}{\emph{Amazon}}.
\newblock
\newblock
\shownote{\url{https://aws.amazon.com/blogs/security/confidential-computing-an-aws-perspective/}}.


\bibitem[IBM(2023)]%
        {IBM-CC}
 \bibinfo{year}{2023}\natexlab{}.
\newblock \showarticletitle{Confidential computing on {IBM Cloud}}. In \bibinfo{booktitle}{\emph{IBM}}.
\newblock
\newblock
\shownote{\url{https://www.ibm.com/cloud/confidential-computing}}.


\bibitem[CoC(2023)]%
        {CoCo}
 \bibinfo{year}{2023}\natexlab{}.
\newblock \showarticletitle{Confidential Containers}.
\newblock
\newblock
\shownote{\url{https://github.com/confidential-containers}}.


\bibitem[Mon(2023)]%
        {Monsoon}
 \bibinfo{year}{2023}\natexlab{}.
\newblock \bibinfo{booktitle}{\emph{High Voltage Power Monitor}}.
\newblock
\newblock
\shownote{\url{https://www.msoon.com/high-voltage-power-monitor}}.


\bibitem[Int(2023)]%
        {Intel-TDX}
 \bibinfo{year}{2023}\natexlab{}.
\newblock \showarticletitle{Intel Trust Domain Extensions}. In \bibinfo{booktitle}{\emph{Intel}}.
\newblock
\newblock
\shownote{\url{https://www.intel.com/content/www/us/en/developer/articles/technical/intel-trust-domain-extensions.html}}.


\bibitem[K8s(2023)]%
        {K8s}
 \bibinfo{year}{2023}\natexlab{}.
\newblock \bibinfo{booktitle}{\emph{Kubernetes}}.
\newblock
\newblock
\shownote{\url{https://kubernetes.io/}}.


\bibitem[lib(2023)]%
        {libjpeg-turbo}
 \bibinfo{year}{2023}\natexlab{}.
\newblock \bibinfo{booktitle}{\emph{libjpeg-turbo}}.
\newblock
\newblock
\shownote{\url{https://libjpeg-turbo.org/}}.


\bibitem[pho(2023)]%
        {photoprism}
 \bibinfo{year}{2023}\natexlab{}.
\newblock \bibinfo{title}{Photoprism}.
\newblock
\newblock
\newblock
\shownote{\url{https://www.photoprism.app/}}.


\bibitem[Red(2023)]%
        {ReduceLROnPlateau}
 \bibinfo{year}{2023}\natexlab{}.
\newblock \showarticletitle{ReduceLROnPlateau}.
\newblock
\newblock
\shownote{\url{https://pytorch.org/docs/stable/generated/\\torch.optim.lr\_scheduler.ReduceLROnPlateau.html}}.


\bibitem[Tru(2023)]%
        {TrustedNetwork}
 \bibinfo{year}{2023}\natexlab{}.
\newblock \bibinfo{title}{Trusted Network}.
\newblock
\newblock
\newblock
\shownote{\url{https://www.sciencedirect.com/topics/computer-science/trusted-network}}.


\bibitem[Vul(2023)]%
        {Vultr}
 \bibinfo{year}{2023}\natexlab{}.
\newblock \bibinfo{title}{Vultr}.
\newblock
\newblock
\newblock
\shownote{\url{https://www.vultr.com}}.


\bibitem[Akter et~al\mbox{.}(2020)]%
        {Akt-Security20}
\bibfield{author}{\bibinfo{person}{Taslima Akter}, \bibinfo{person}{Bryan Dosono}, \bibinfo{person}{Tousif Ahmed}, \bibinfo{person}{Apu Kapadia}, {and} \bibinfo{person}{Bryan Semaan}.} \bibinfo{year}{2020}\natexlab{}.
\newblock \showarticletitle{" I am uncomfortable sharing what I can't see": Privacy Concerns of the Visually Impaired with Camera Based Assistive Applications}. In \bibinfo{booktitle}{\emph{29th USENIX Security Symposium (USENIX Security 20)}}. \bibinfo{pages}{1929--1948}.
\newblock


\bibitem[Alhilal et~al\mbox{.}(2022)]%
        {Alh-WWW22}
\bibfield{author}{\bibinfo{person}{Ahmad Alhilal}, \bibinfo{person}{Tristan Braud}, \bibinfo{person}{Bo Han}, {and} \bibinfo{person}{Pan Hui}.} \bibinfo{year}{2022}\natexlab{}.
\newblock \showarticletitle{Nebula: Reliable low-latency video transmission for mobile cloud gaming}. In \bibinfo{booktitle}{\emph{ACM Web Conference 2022}}. \bibinfo{pages}{3407--3417}.
\newblock


\bibitem[Ball{\'e} et~al\mbox{.}(2015)]%
        {Balle-arXiv15}
\bibfield{author}{\bibinfo{person}{Johannes Ball{\'e}}, \bibinfo{person}{Valero Laparra}, {and} \bibinfo{person}{Eero~P Simoncelli}.} \bibinfo{year}{2015}\natexlab{}.
\newblock \showarticletitle{Density modeling of images using a generalized normalization transformation}.
\newblock \bibinfo{journal}{\emph{arXiv}} (\bibinfo{year}{2015}).
\newblock


\bibitem[Ball{\'e} et~al\mbox{.}(2018)]%
        {Balle-arXiv18}
\bibfield{author}{\bibinfo{person}{Johannes Ball{\'e}}, \bibinfo{person}{David Minnen}, \bibinfo{person}{Saurabh Singh}, \bibinfo{person}{Sung~Jin Hwang}, {and} \bibinfo{person}{Nick Johnston}.} \bibinfo{year}{2018}\natexlab{}.
\newblock \showarticletitle{Variational image compression with a scale hyperprior}.
\newblock \bibinfo{journal}{\emph{arXiv}} (\bibinfo{year}{2018}).
\newblock


\bibitem[B{\'e}gaint et~al\mbox{.}(2020)]%
        {Begaint-arXiv20}
\bibfield{author}{\bibinfo{person}{Jean B{\'e}gaint}, \bibinfo{person}{Fabien Racap{\'e}}, \bibinfo{person}{Simon Feltman}, {and} \bibinfo{person}{Akshay Pushparaja}.} \bibinfo{year}{2020}\natexlab{}.
\newblock \showarticletitle{CompressAI: a PyTorch library and evaluation platform for end-to-end compression research}.
\newblock \bibinfo{journal}{\emph{arXiv preprint arXiv:2011.03029}} (\bibinfo{year}{2020}).
\newblock


\bibitem[Cerqueira et~al\mbox{.}(2014)]%
        {Cer-ICNC14}
\bibfield{author}{\bibinfo{person}{Eduardo Cerqueira}, \bibinfo{person}{Euisin Lee}, \bibinfo{person}{Jui-Ting Weng}, \bibinfo{person}{Jae-Han Lim}, \bibinfo{person}{Joshua Joy}, {and} \bibinfo{person}{Mario Gerla}.} \bibinfo{year}{2014}\natexlab{}.
\newblock \showarticletitle{Recent advances and challenges in human-centric multimedia mobile cloud computing}. In \bibinfo{booktitle}{\emph{International Conference on Computing, Networking and Communications (ICNC)}}. \bibinfo{pages}{242--246}.
\newblock


\bibitem[Chattopadhayay and Rijal(2023)]%
        {Cha-CCWC23}
\bibfield{author}{\bibinfo{person}{Ankur Chattopadhayay} {and} \bibinfo{person}{Isha Rijal}.} \bibinfo{year}{2023}\natexlab{}.
\newblock \showarticletitle{Towards Inclusive Privacy Consenting for GDPR Compliance in Visual Surveillance: A Survey Study}. In \bibinfo{booktitle}{\emph{2023 IEEE 13th Annual Computing and Communication Workshop and Conference (CCWC)}}. \bibinfo{pages}{1287--1293}.
\newblock


\bibitem[Chen et~al\mbox{.}(2017)]%
        {Chen-VehicularComm17}
\bibfield{author}{\bibinfo{person}{Chien-Hung Chen}, \bibinfo{person}{Che-Rung Lee}, {and} \bibinfo{person}{Walter Chen-Hua Lu}.} \bibinfo{year}{2017}\natexlab{}.
\newblock \showarticletitle{Smart in-car camera system using mobile cloud computing framework for deep learning}.
\newblock \bibinfo{journal}{\emph{Vehicular Communications}}  \bibinfo{volume}{10} (\bibinfo{year}{2017}), \bibinfo{pages}{84--90}.
\newblock


\bibitem[Clark(2015)]%
        {pillow}
\bibfield{author}{\bibinfo{person}{Alex Clark}.} \bibinfo{year}{2015}\natexlab{}.
\newblock \bibinfo{title}{Pillow (PIL Fork) Documentation}.
\newblock
\newblock
\urldef\tempurl%
\url{https://buildmedia.readthedocs.org/media/pdf/pillow/latest/pillow.pdf}
\showURL{%
\tempurl}


\bibitem[Khan et~al\mbox{.}(2022)]%
        {Khan-Access22}
\bibfield{author}{\bibinfo{person}{Muhammad~Asif Khan}, \bibinfo{person}{Emna Baccour}, \bibinfo{person}{Zina Chkirbene}, \bibinfo{person}{Aiman Erbad}, \bibinfo{person}{Ridha Hamila}, \bibinfo{person}{Mounir Hamdi}, {and} \bibinfo{person}{Moncef Gabbouj}.} \bibinfo{year}{2022}\natexlab{}.
\newblock \showarticletitle{A survey on mobile edge computing for video streaming: Opportunities and challenges}.
\newblock \bibinfo{journal}{\emph{IEEE Access}} (\bibinfo{year}{2022}).
\newblock


\bibitem[Kingma and Ba(2014)]%
        {Kingma-arXiv14}
\bibfield{author}{\bibinfo{person}{Diederik~P Kingma} {and} \bibinfo{person}{Jimmy Ba}.} \bibinfo{year}{2014}\natexlab{}.
\newblock \showarticletitle{Adam: A method for stochastic optimization}.
\newblock \bibinfo{journal}{\emph{arXiv}} (\bibinfo{year}{2014}).
\newblock


\bibitem[Krizhevsky et~al\mbox{.}(2009)]%
        {cifar10}
\bibfield{author}{\bibinfo{person}{Alex Krizhevsky}, \bibinfo{person}{Geoffrey Hinton}, {et~al\mbox{.}}} \bibinfo{year}{2009}\natexlab{}.
\newblock \showarticletitle{Learning multiple layers of features from tiny images}.
\newblock  (\bibinfo{year}{2009}).
\newblock


\bibitem[Padilla-L{\'o}pez et~al\mbox{.}(2015)]%
        {Pad-ESA15}
\bibfield{author}{\bibinfo{person}{Jos{\'e}~Ram{\'o}n Padilla-L{\'o}pez}, \bibinfo{person}{Alexandros~Andre Chaaraoui}, {and} \bibinfo{person}{Francisco Fl{\'o}rez-Revuelta}.} \bibinfo{year}{2015}\natexlab{}.
\newblock \showarticletitle{Visual privacy protection methods: A survey}.
\newblock \bibinfo{journal}{\emph{Expert Systems with Applications}} \bibinfo{volume}{42}, \bibinfo{number}{9} (\bibinfo{year}{2015}), \bibinfo{pages}{4177--4195}.
\newblock


\bibitem[Pletcher(2023)]%
        {Ple-arxiv23}
\bibfield{author}{\bibinfo{person}{Scott Pletcher}.} \bibinfo{year}{2023}\natexlab{}.
\newblock \showarticletitle{Visual Privacy: Current and Emerging Regulations Around Unconsented Video Analytics in Retail}.
\newblock \bibinfo{journal}{\emph{arXiv preprint arXiv:2302.12935}} (\bibinfo{year}{2023}).
\newblock


\bibitem[Ren et~al\mbox{.}(2018)]%
        {Ren-ECCV18}
\bibfield{author}{\bibinfo{person}{Zhongzheng Ren}, \bibinfo{person}{Yong~Jae Lee}, {and} \bibinfo{person}{Michael~S Ryoo}.} \bibinfo{year}{2018}\natexlab{}.
\newblock \showarticletitle{Learning to anonymize faces for privacy preserving action detection}. In \bibinfo{booktitle}{\emph{European conference on computer vision (ECCV)}}. \bibinfo{pages}{620--636}.
\newblock


\bibitem[Stangl et~al\mbox{.}(2022)]%
        {Sta-TACCESS22}
\bibfield{author}{\bibinfo{person}{Abigale Stangl}, \bibinfo{person}{Kristina Shiroma}, \bibinfo{person}{Nathan Davis}, \bibinfo{person}{Bo Xie}, \bibinfo{person}{Kenneth~R Fleischmann}, \bibinfo{person}{Leah Findlater}, {and} \bibinfo{person}{Danna Gurari}.} \bibinfo{year}{2022}\natexlab{}.
\newblock \showarticletitle{Privacy concerns for visual assistance technologies}.
\newblock \bibinfo{journal}{\emph{ACM Transactions on Accessible Computing (TACCESS)}} \bibinfo{volume}{15}, \bibinfo{number}{2} (\bibinfo{year}{2022}), \bibinfo{pages}{1--43}.
\newblock


\bibitem[Sun et~al\mbox{.}(2018)]%
        {Sun-CVPR18}
\bibfield{author}{\bibinfo{person}{Qianru Sun}, \bibinfo{person}{Liqian Ma}, \bibinfo{person}{Seong~Joon Oh}, \bibinfo{person}{Luc Van~Gool}, \bibinfo{person}{Bernt Schiele}, {and} \bibinfo{person}{Mario Fritz}.} \bibinfo{year}{2018}\natexlab{}.
\newblock \showarticletitle{Natural and effective obfuscation by head inpainting}. In \bibinfo{booktitle}{\emph{IEEE Conference on Computer Vision and Pattern Recognition (CVPR)}}. \bibinfo{pages}{5050--5059}.
\newblock


\bibitem[Tian et~al\mbox{.}(2017)]%
        {Tian-CNS17}
\bibfield{author}{\bibinfo{person}{Yifan Tian}, \bibinfo{person}{Yantian Hou}, {and} \bibinfo{person}{Jiawei Yuan}.} \bibinfo{year}{2017}\natexlab{}.
\newblock \showarticletitle{Capia: Cloud assisted privacy-preserving image annotation}. In \bibinfo{booktitle}{\emph{2017 IEEE Conference on Communications and Network Security (CNS)}}. \bibinfo{pages}{1--9}.
\newblock


\bibitem[Wang and Dey(2013)]%
        {Wang-TM13}
\bibfield{author}{\bibinfo{person}{Shaoxuan Wang} {and} \bibinfo{person}{Sujit Dey}.} \bibinfo{year}{2013}\natexlab{}.
\newblock \showarticletitle{Adaptive mobile cloud computing to enable rich mobile multimedia applications}.
\newblock \bibinfo{journal}{\emph{IEEE Transactions on Multimedia}} \bibinfo{volume}{15}, \bibinfo{number}{4} (\bibinfo{year}{2013}), \bibinfo{pages}{870--883}.
\newblock


\bibitem[Wu et~al\mbox{.}(2021)]%
        {Wu-MobiCom21}
\bibfield{author}{\bibinfo{person}{Hao Wu}, \bibinfo{person}{Xuejin Tian}, \bibinfo{person}{Minghao Li}, \bibinfo{person}{Yunxin Liu}, \bibinfo{person}{Ganesh Ananthanarayanan}, \bibinfo{person}{Fengyuan Xu}, {and} \bibinfo{person}{Sheng Zhong}.} \bibinfo{year}{2021}\natexlab{}.
\newblock \showarticletitle{PECAM: privacy-enhanced video streaming and analytics via securely-reversible transformation}. In \bibinfo{booktitle}{\emph{Annual International Conference on Mobile Computing and Networking}}. \bibinfo{pages}{229--241}.
\newblock


\bibitem[Xu and Mao(2013)]%
        {Xu-WirelessComm13}
\bibfield{author}{\bibinfo{person}{Yi Xu} {and} \bibinfo{person}{Shiwen Mao}.} \bibinfo{year}{2013}\natexlab{}.
\newblock \showarticletitle{A survey of mobile cloud computing for rich media applications}.
\newblock \bibinfo{journal}{\emph{IEEE Wireless Communications}} \bibinfo{volume}{20}, \bibinfo{number}{3} (\bibinfo{year}{2013}), \bibinfo{pages}{46--53}.
\newblock


\bibitem[Yang et~al\mbox{.}(2023)]%
        {Yang-FTCGV23}
\bibfield{author}{\bibinfo{person}{Yibo Yang}, \bibinfo{person}{Stephan Mandt}, \bibinfo{person}{Lucas Theis}, {et~al\mbox{.}}} \bibinfo{year}{2023}\natexlab{}.
\newblock \showarticletitle{An introduction to neural data compression}.
\newblock \bibinfo{journal}{\emph{Foundations and Trends{\textregistered} in Computer Graphics and Vision}} \bibinfo{volume}{15}, \bibinfo{number}{2} (\bibinfo{year}{2023}), \bibinfo{pages}{113--200}.
\newblock


\bibitem[Ye et~al\mbox{.}(2022)]%
        {Ye-MMSys22}
\bibfield{author}{\bibinfo{person}{Mengmei Ye}, \bibinfo{person}{Zhongze Tang}, \bibinfo{person}{Huy Phan}, \bibinfo{person}{Yi Xie}, \bibinfo{person}{Bo Yuan}, {and} \bibinfo{person}{Sheng Wei}.} \bibinfo{year}{2022}\natexlab{}.
\newblock \showarticletitle{Visual privacy protection in mobile image recognition using protective perturbation}. In \bibinfo{booktitle}{\emph{ACM Multimedia Systems Conference (MMSys)}}. \bibinfo{pages}{164--176}.
\newblock


\bibitem[Zhao et~al\mbox{.}(2023)]%
        {Zhao-arxiv23}
\bibfield{author}{\bibinfo{person}{Ruoyu Zhao}, \bibinfo{person}{Yushu Zhang}, \bibinfo{person}{Tao Wang}, \bibinfo{person}{Wenying Wen}, \bibinfo{person}{Yong Xiang}, {and} \bibinfo{person}{Xiaochun Cao}.} \bibinfo{year}{2023}\natexlab{}.
\newblock \showarticletitle{Visual Content Privacy Protection: A Survey}.
\newblock \bibinfo{journal}{\emph{arXiv preprint arXiv:2303.16552}} (\bibinfo{year}{2023}).
\newblock


\bibitem[Zhao et~al\mbox{.}(2022)]%
        {Zhao-TMC22}
\bibfield{author}{\bibinfo{person}{Yi Zhao}, \bibinfo{person}{Zheng Yang}, \bibinfo{person}{Xiaowu He}, \bibinfo{person}{Xinjun Cai}, \bibinfo{person}{Xin Miao}, {and} \bibinfo{person}{Qiang Ma}.} \bibinfo{year}{2022}\natexlab{}.
\newblock \showarticletitle{Trine: Cloud-edge-device cooperated real-time video analysis for household applications}.
\newblock \bibinfo{journal}{\emph{IEEE Transactions on Mobile Computing}} (\bibinfo{year}{2022}).
\newblock


\bibitem[Zhu et~al\mbox{.}(2020)]%
        {Zhu-AIES20}
\bibfield{author}{\bibinfo{person}{Bingquan Zhu}, \bibinfo{person}{Hao Fang}, \bibinfo{person}{Yanan Sui}, {and} \bibinfo{person}{Luming Li}.} \bibinfo{year}{2020}\natexlab{}.
\newblock \showarticletitle{Deepfakes for Medical Video De-Identification: Privacy Protection and Diagnostic Information Preservation}. In \bibinfo{booktitle}{\emph{AAAI/ACM Conference on AI, Ethics, and Society (AIES)}}. \bibinfo{pages}{414--420}.
\newblock


\end{thebibliography}
\end{document}